\documentclass[prc,floatfix,groupedaddress,nofootinbib,showpacs,preprintnumbers,
amsmath,amssymb,amsfonts,superscriptaddress,widetable] {revtex4}
\usepackage{graphicx}
\usepackage{dcolumn}
\usepackage{mathrsfs}
\usepackage{bm}

\usepackage{graphicx}
\usepackage{dcolumn}
\usepackage{bm}
\usepackage[usenames]{color}
\begin{document}

\title{Giant monopole energies from a constrained relativistic mean-field approach}
\author{Wei-Chia Chen}
\email{wc09c@my.fsu.edu} 
\affiliation{Department of Physics, Florida State University, Tallahassee, FL 32306} 
\author{J. Piekarewicz}
\email{jpiekarewicz@fsu.edu}
\affiliation{Department of Physics, Florida State University, Tallahassee, FL 32306}
\author{M. Centelles}
\email{mariocentelles@ub.edu}
\affiliation{Departament d'Estructura i Constituents de la Mat\`eria
 and Institut de Ci\`encies del Cosmos, Facultat de F\'{\i}sica,
 Universitat de Barcelona, Diagonal {\sl 645}, {\sl 08028} Barcelona, Spain} 
\date{\today}
\begin{abstract}
\begin{description}
 \item[Background:] Average energies of nuclear collective
  modes may be efficiently and accurately computed using 
  a non-relativistic constrained approach without
  reliance on a random phase approximation (RPA).
 \item[Purpose:] To extend the constrained approach to the
 relativistic domain and to establish its impact on
 the calibration of energy density functionals.
 \item[Methods:] Relativistic RPA calculations of the giant 
 monopole resonance (GMR) are compared against the 
 predictions of the corresponding constrained approach 
 using two accurately calibrated energy density functionals. 
 \item[Results:] We find excellent agreement---at the 2\% level
 or better---between the predictions of the relativistic RPA and 
 the corresponding constrained approach for magic (or 
 semi-magic) nuclei ranging from ${}^{16}$O to ${}^{208}$Pb.
 \item[Conclusions:] An efficient and accurate method is
 proposed for incorporating nuclear collective excitations 
 into the calibration of future energy density functionals.
\end{description}
\end{abstract}
\pacs{21.60.Jz, 21.65.Cd, 21.65.Mn} 
\maketitle

\section{Introduction}
\label{intro}
A tractable microscopic theory that both predicts and provides
well-quantified theoretical uncertainties for nuclear properties
throughout the nuclear chart is the guiding principle in the quest of a
{\sl Universal Nuclear Energy Density
Functional}~\cite{UNEDF,Kortelainen:2010hv}. {\sl Density functional
theory} (DFT) provides a powerful---and perhaps unique---framework for
the accurate calculation of the ground-state properties and collective
excitations of medium-to-heavy nuclei. Based on the seminal work by
Kohn and collaborators~\cite{Hohenberg:1964zz,Kohn:1965,Kohn:1999},
DFT shifts the focus from the complicated many-body wave function to
the much simpler one-body density.  By doing so, the formidable
challenge of deducing the exact ground-state energy and one-body
density from the many-body wave function {\sl ``reduces''} to the
minimization of a suitable functional of the density. Of course, this
enormous simplification comes at a price. Whereas DFT establishes the
existence of a well-defined density functional, it offers no guidance
on how to build it.  Nevertheless, by strictly focusing on
measurable quantities, DFT offers the opportunity for using physical
intuition and symmetries/constraints to guide the construction of the
functional. Moreover, Kohn and Sham have shown how the ground-state
energy and corresponding one-body density may be obtained from a
variational problem that reduces to a self-consistent solution of a
set of mean-field-like ({\sl ``Kohn-Sham''})
equations~\cite{Kohn:1965}. However, although the form of the
Kohn-Sham equations is reminiscent of the self-consistent Hartree
(or Hartree-Fock) problem in the presence of an underlying (bare) 
nucleon-nucleon (NN) interaction, the constants that parametrize the 
Kohn-Sham potential are directly fitted to many-body properties (such as 
masses and charge radii) rather than two-body data.  In this manner the 
complicated many-body dynamics gets implicitly encoded in the parameters 
of the model.  In principle, a proper implementation of DFT and the Kohn-Sham
equations incorporates {\sl all many-body effects} in quantities that are 
functionals of the one-body density, such as the total ground-state 
energy\,\cite{Kohn:1999}.

The construction of an energy density functional (EDF) starts with the
selection of terms that incorporate important symmetries and features
of the nuclear dynamics. Although in principle such terms may be
inspired by the underlying nucleon-nucleon dynamics, the explicit
value of the coefficients in front of these terms ({\sl i.e.,} the
model parameters) is customarily obtained through the minimization 
of a quality ({\sl e.g.,} $\chi^{2}$) function. Thus, the model
parameters may---and in general do---differ significantly from the
{\sl ``bare''} NN values. This is a reflection of the fact that the parameters of
the model are calibrated to physical observables that incorporate few-
and many-body correlations. In the language of DFT, most of the
complicated many-body dynamics gets implicitly encoded in the
parameters of the model.  Once the EDF has been properly defined, a
set of accurately-measured ground-state observables is selected to
constrain the model parameters through the minimization of a quality
function. However, given that ground-state observables are fairly
insensitive to fluctuations around the equilibrium density, one must
often resort to {\sl ``pseudo-data''}---in the form of various bulk
properties of infinite nuclear matter---to better constrain the
functional. Regardless, once the functional is properly defined and a
fitting protocol established, minimization of the quality function
defines the model. Until recently, once the minimum was found, one
proceeded to validate the model against observables not included in
the quality function~\cite{Piekarewicz:2007dx}. Lately, however, a few
studies have been devoted to map the landscape around the
$\chi^{2}$-minimum~\cite{Reinhard:2010wz, Fattoyev:2011ns,
Fattoyev:2012rm}. Among the wealth of information revealed by such
detailed statistical studies is the degree of correlation among
various observables. Moreover, one can also use such a covariance
analysis to assess the robustness of the model.  For example, one
could ask whether certain linear combinations of model parameters
remain poorly constrained by the choice of observables.  We find this
to be particularly true in the case of the isovector sector that is
hindered by the unavailability of highly accurate data on neutron
skins~\cite{Fattoyev:2011ns}.

It is the main goal of the present contribution to explore the
feasibility of supplementing ground-state observables with nuclear
excitations in the calibration of the quality function of relativistic functionals.
We are confident that such an approach will relax the need for
pseudo-data while providing better constraints on the determination of
the model parameters. Specifically, we advocate supplementing the
quality function with centroid energies of monopole resonances for
nuclei with a wide range of neutron-proton asymmetries. By doing so,
the quality function becomes highly sensitive to the incompressibility
of neutron-rich matter---which itself depends on both the compression
modulus of symmetric matter and the slope of the symmetry
energy\,\cite{Piekarewicz:2003br,Piekarewicz:2008nh,Piekarewicz:2009gb}.

Perhaps the most serious impediment to the implementation of this
program is numerical in nature. By itself, computing the distribution
of monopole strength for a wide range of excitation energies for
models with finite-range interactions is numerically
intensive\,\cite{Piekarewicz:2001nm}. Thus, embedding such RPA
calculation into a complex $\chi^{2}$-minimization routine becomes
impractical even for today's most powerful computers. However, the
computation of centroid energies relies on knowledge of just a few
moments of the distribution.  In particular, an estimate of the
centroid energy may be obtained from knowledge of two moments: the
energy weighted sum ($m_{1}$) and the {\sl inverse} energy weighted 
sum ($m_{-1}$). Remarkably, for non-relativistic density functionals it 
has since long been established that the fully self-consistent $m_{-1}$ 
moment may be calculated quite generally from the ground-state properties 
of a slightly modified ({\sl i.e., ``constrained''}) density 
functional\,\cite{Thouless:1960,Thouless:1961,Bohigas:1978qu,
Marshalek:1973,Ring:2004}, a result often referred to as the 
{\sl ``dielectric theorem''}. Similarly, the $m_{1}$ moment may also be 
obtained by computing the ground-state expectation value of a suitable 
operator ($\langle r^{2}\rangle$ in the case of the isoscalar monopole 
mode)\,\cite{Harakeh:2001}. Indeed, the constrained approach has been 
widely used in the non-relativistic mean-field theory to study the centroid
energy of the giant monopole resonance on a variety of nuclei, see 
Refs.\,\cite{Colo:2004mj,Sil:2006sh,Bohigas:1978qu,Gleissl:1989vh,Blaizot:1995, 
Centelles:2005fg,Capelli:2009zz,Khan:2009,Khan:2010mv} and references quoted
therein. In particular, in response to significant experimental advances that have 
allowed the measurement of the distribution of monopole strength in neutron-rich 
nuclei\,\cite{Li:2007bp,Patel:2012}, constrained Skyrme calculations have been
used to predict the GMR centroid energies along different isotopic 
chains\,\cite{Centelles:2005fg,Capelli:2009zz,Khan:2009}. Measurements of 
the distribution of monopole strength along isotopic chains are sensitive to 
the incompressibility coefficient of asymmetric matter, and thus to the poorly 
known density dependence of the nuclear symmetry energy\,\cite{Piekarewicz:2008nh}.
Therefore, for the systematic numerical exploration required to elucidate
the isovector sector of the EDF, the constrained approach dramatically 
simplifies the computational effort as compared with the numerically 
intensive demands of self-consistent RPA calculations.

In contrast to the non-relativistic case, only a handful of constrained calculations 
have been reported in the context of relativistic mean-field theory. As far as we 
know, the first relativistic constrained Hartree calculations of giant monopole 
energies of finite nuclei were performed at the end of the 1980s using the linear 
Walecka model\,\cite{Maruyama:1988hh,Boersma:1991uj}. Some time later, 
constrained monopole energies using non-linear RMF models were reported in 
Ref.\,\cite{Stoitsov:1994ne}. More recently, constrained RMF calculations of the 
$m_{-1}$ moment of the monopole mode were carried out in Ref.~\cite{Ma:2001}. 
It is also worth mentioning that a somewhat different constrained approach, the 
so-called generator coordinate method (GCM), has been applied in RMF studies 
of giant monopole resonances\,\cite{Stoitsov:1994bj,Vretenar:1997,Sharma:2009}. 
However, the available literature with quantitative analyses of the degree of agreement 
between relativistic constrained calculations and the corresponding RPA results is 
very scarce; we are aware of only the RMF constrained-vs-RPA study of the 
$m_{-1}$ moment of the GMR in ${}^{208}$Pb reported in Ref.\,\cite{Ma:2001} . 
Moreover, to our knowledge the constrained approach has not been systematically 
implemented in the context of relativistic mean-field theories with the goal of 
supplementing the calibration of nuclear functionals with collective excitations. 
Our work aims at filling this gap in the literature.

The reason for such an imbalance between the constrained relativistic and
non-relativistic frameworks may be fairly evident. On the one hand, the
derivation of the energy weighted sum rule (EWSR) hinges on a
non-relativistic kinetic energy operator (i.e., quadratic in the
momentum), and on the other hand, the derivation of the inverse energy
weighted sum rule through the dielectric theorem has been established
only for non-relativistic Hamiltonians. Thus, whereas the sum-rule
theorems have been proven analytically in the non-relativistic case,
to our knowledge general proofs do not exist in the relativistic RPA 
theory (except for a suggestion in Sec.\,3 of Ref.\,\cite{Ma:2001} 
that the relativistic RPA $m_{-1}$ moment can be obtained from a 
constrained RMF calculation). Nevertheless, extending the 
constrained approach to the relativistic domain seems plausible 
by the fact that accurately calibrated non-relativistic and relativistic 
energy density functionals provide similar distributions of isoscalar 
monopole strength. In particular, DFT makes no demands on whether 
the nuclear functional should be of a relativistic or non-relativistic 
character; the precise form of the functional then becomes a matter 
of convenience. It may also be mentioned that in recent constrained 
RMF calculations performed in the Thomas-Fermi approximation a 
close analogy of the results with the classical sum-rule expressions 
was reported\,\cite{Centelles:2010}.

In the present paper we perform a numerical study to explore the quality 
of the constrained RMF predictions against detailed relativistic RPA
calculations. Inspired by the appeal of the constrained approach in the 
non-relativistic case---and the absence (to our knowledge) of detailed 
proofs of relativistic sum-rule theorems---we consider the present 
numerical validation of the constrained approach in the relativistic domain 
both interesting and important. We also note that correlating GMR energies
to the bulk properties of the equation of state (EOS) is critical in our 
quest for imposing meaningful constraints on the nuclear EOS at and below 
saturation density. To be able to draw general conclusions on such
correlations, they should be systematically investigated using a 
variety of nuclear energy density functionals, both non-relativistic and
relativistic. Indeed, constrained calculations of the GMR energy are 
being used for exactly those reasons\,\cite{Khan:2012}. It thus seems 
timely and necessary to establish the accuracy of the constrained RMF 
predictions of giant monopole energies. 

The manuscript has been organized as follows. In Sec.~\ref{Formalism}
we review the necessary formalism required to compute the excitation
energy from the complete distribution of isoscalar monopole strength.
In addition, we describe how the $m_{-1}$ and $m_{1}$ moments are
computed in the constrained approach. In Sec.~\ref{Results} we compare
results obtained from the constrained approach against those extracted
from the full distribution of strength. We conclude in Sec.~\ref{Conclusions}
with a summary of our results and plans for the future.

\section{Formalism}
\label{Formalism}

The starting point for the calculation of the nuclear response is the
interacting Lagrangian density of Ref.~\cite{Mueller:1996pm}
supplemented by an isoscalar-isovector term first introduced 
in Ref.~\cite{Horowitz:2000xj}. That is,
\begin{eqnarray}
{\mathscr L}_{\rm int} &=&
\bar\psi \left[g_{\rm s}\phi   \!-\! 
         \left(g_{\rm v}V_\mu  \!+\!
    \frac{g_{\rho}}{2}{\mbox{\boldmath $\tau$}}\cdot{\bf b}_{\mu} 
                               \!+\!    
    \frac{e}{2}(1\!+\!\tau_{3})A_{\mu}\right)\gamma^{\mu}
         \right]\psi \nonumber \\
                   &-& 
    \frac{\kappa}{3!} (g_{\rm s}\phi)^3 \!-\!
    \frac{\lambda}{4!}(g_{\rm s}\phi)^4 \!+\!
    \frac{\zeta}{4!}   g_{\rm v}^4(V_{\mu}V^\mu)^2 +
   \Lambda_{\rm v}\Big(g_{\rho}^{2}\,{\bf b}_{\mu}\cdot{\bf b}^{\mu}\Big)
                           \Big(g_{\rm v}^{2}V_{\nu}V^{\nu}\Big)\;.
 \label{LDensity}
\end{eqnarray}
The Lagrangian density includes an isodoublet nucleon field ($\psi$)
interacting via the exchange of two isoscalar mesons, a scalar
($\phi$) and a vector ($V^{\mu}$), one isovector meson ($b^{\mu}$),
and the photon ($A^{\mu}$)~\cite{Serot:1984ey,Serot:1997xg}. In
addition to meson-nucleon interactions, the Lagrangian density is
supplemented by four nonlinear meson interactions with coupling
constants denoted by $\kappa$, $\lambda$, $\zeta$, and $\Lambda_{\rm v}$.  
The first two terms ($\kappa$ and $\lambda$) are responsible for
a softening of the equation of state of symmetric nuclear matter at
normal density that results in a significant reduction of the
incompressibility coefficient relative to that of the original Walecka
model~\cite{Walecka:1974qa,Boguta:1977xi,Serot:1984ey}. Indeed,
such a softening is demanded by the measured distribution of isoscalar
monopole strength in medium to heavy
nuclei~\cite{Youngblood:1999,Uchida:2003,Uchida:2004bs,Li:2007bp,
Li:2010kfa,Piekarewicz:2001nm,Piekarewicz:2002jd,Colo:2004mj}. 
Further, omega-meson self-interactions, as described
by the parameter $\zeta$, serve to soften the equation of state of
symmetric nuclear matter at high densities and at present can 
only be meaningfully constrained by the limiting masses of neutron
stars~\cite{Demorest:2010bx}.  Finally, the parameter 
$\Lambda_{\rm v}$ induces isoscalar-isovector mixing and is
responsible for modifying the poorly-constrained density dependence of
the symmetry energy~\cite{Horowitz:2000xj,Horowitz:2001ya}.  Tuning
this parameter has served to uncover correlations between the neutron
skin of a heavy nucleus---such as ${}^{208}$Pb---and a host of both
laboratory and astrophysical observables.

A consistent mean-field plus RPA (MF+RPA) approach to the nuclear
response starts with the calculation of various ground-state
properties. This procedure is implemented by solving self-consistently
the appropriate set of mean-field ({\sl i.e.,} Kohn-Sham) equations
generated by the Lagrangian density given above~\cite{Serot:1984ey}.
For the various meson fields one must solve Klein-Gordon equations
with the appropriate baryon densities appearing as their source
terms. These baryon densities are computed from the nucleon orbitals
that are, in turn, obtained from solving the one-body Dirac equation
in the presence of scalar and time-like vector potentials.  This
procedure must then be repeated until self-consistency is achieved.
What emerges from such a self-consistent procedure is a set of
single-particle energies, a corresponding set of Dirac orbitals,
scalar and time-like vector mean-field potentials, and ground-state
densities. A detailed implementation of this procedure may be found 
in Ref.~\cite{Todd:2003xs}.
Having computed various ground-state properties one is now in a
position to compute the linear response of the mean-field ground state
to a variety of probes.  In the present case we are interested in
computing the isoscalar monopole response as probed, for example, in
$\alpha$-scattering experiments\,\cite{Youngblood:1999,Uchida:2003,
Uchida:2004bs,Li:2007bp,Li:2010kfa}.
Although the MF+RPA calculations carried out here follow closely the
formalism developed in much greater detail in
Ref.~\cite{Piekarewicz:2001nm}, some essential details are repeated
here for completeness. 

The distribution of isoscalar monopole strength may be extracted from
the imaginary part of a suitable polarization tensor that we compute
in a consistent relativistic RPA
approximation~\cite{Fetter:1971,Dickhoff:2005}.
That is,
\begin{equation}
  S_{L}(q,\omega)=\sum_{n}\Big|\langle\Psi_{n}|\hat{\rho}({\bf q})|
  \Psi_{0}\rangle\Big|^{2}\delta(\omega-\omega_{n}) =-\frac{1}{\pi}
   \Im\,\Pi^{00}_{\rm RPA}({\bf q},{\bf q};\omega) \;,
 \label{S00}
\end{equation}
where $\Psi_{0}$ is the nuclear ground state and $\Psi_{n}$ is an
excited state with excitation energy $\omega_{n}\!=\!E_{n}\!-\!E_{0}$.
To excite isoscalar monopole modes a simple transition operator of 
the following form may be used:
\begin{equation}
 \hat{\rho}({\bf q}) \!=\! \int d^{3}r \, \bar{\psi}({\bf r}) 
  e^{-i{\bf q}\cdot{\bf r}} \gamma^{0}\psi({\bf r}) \;.
 \label{Rhoq}
\end{equation}
Here $\hat{\rho}({\bf q})$ is the Fourier transform of the isoscalar
(baryon) density and $\gamma^{0}\!=\!{\rm diag}(1,1,-1,-1)$ is the
zeroth (or timelike) component of the Dirac matrices.  Such a
transition operator is capable of exciting all natural-parity states,
including the isoscalar monopole ($E0$) mode of
interest to this work. In the particular case of an $E0$ ($J^{\pi}\!=\!0^{+}$)
excitation, the effective transition operator reduces to the following
simple form:
\begin{equation}
  \hat{\rho}_{_{E0}}({\bf q}) = \int d^{3}r \,\bar{\psi}({\bf r}) j_{0}(qr) 
  \gamma^{0}\psi({\bf r}) \;,
 \label{RhoGMR}
\end{equation}
where $j_{0}(x)\!=\!\sin(x)/x$ is the spherical Bessel function of
order zero. Finally, in the long-wavelength limit, the distribution 
of isoscalar monopole strength $R(\omega;E0)$ may be directly 
extracted from the longitudinal response. That is,
\begin{equation}
  R(\omega;E0) = \sum_{n}\Big|\langle\Psi_{n}|\hat{r}^{2}|
  \Psi_{0}\rangle\Big|^{2}\delta(\omega-\omega_{n}) =
  \lim_{q\rightarrow 0} \left(\frac{36}{q^{4}}\right) S_{L}(q,\omega;E0) \;. 
  \label{RGMR}
\end{equation}
Connecting the nuclear response to the polarization tensor is highly
appealing as one can bring to bear the full power of the many-body
formalism into the calculation of observables than can be directly
extracted from experiment~\cite{Fetter:1971,Dickhoff:2005}.  Moreover,
the spectral content of the polarization tensor is both illuminating
and physically intuitive. However, enforcing the self-consistency of
the formalism, while essential, is highly non-trivial.  Yet maintaining
self-consistency---in particular by using a residual particle-hole
interaction that is identical to the one used to generate the
mean-field ground state---is essential for the preservation of
important symmetries and the decoupling of various spurious
modes~\cite{Dawson:1990wp,Piekarewicz:2001nm}. Finally,
given the critical role that certain moments of the distribution of
strength play in our discussion, we close this section with a few
essential definitions and relations.

In general, the moments of the distribution of isoscalar-monopole 
strength are defined as follows:
\begin{equation}    
  m_{n}(E0) \equiv \int_{0}^{\infty}\!\omega^{n} R(\omega;E0)\, d\omega \;. 
 \label{Moments}
\end{equation}
In particular, the EWSR for the isoscalar monopole mode is given, for a 
non-relativistic Hamiltonian, by\,\cite{Harakeh:2001}
\begin{equation}    
  m_{1}(E0) = \int_{0}^{\infty}\!\omega R(\omega;E0)\,d\omega 
  = \frac{2\hbar^2}{M} A\langle r^2\rangle
  \equiv \frac{2\hbar^2}{M} \int r^{2}\rho({\bf r})\,d^{3}r \;,
 \label{EWSR}
\end{equation}
where $M$ is the nucleon mass and $\rho({\bf r})$ is the ground-state baryon 
density normalized to the baryon number $A$. In essence,
the power of the sum rule is that it relates a moment of the full RPA
distribution to the mean-square value of the ground-state density.  
We note that the above {\sl classical sum rule} is only valid in
the absence of exchange and momentum-dependent
forces~\cite{Harakeh:2001}.  Such forces modify the classical sum
rules and their impact is traditionally accounted for by multiplying
the right-hand side of Eq.~(\ref{EWSR}) by a correction factor.
Perhaps the best known case for the need of such a correction factor
($\kappa$) is in the context of the photo-absorption cross
section and the model-independent Thomas-Reiche-Kuhn sum rule (with
$\kappa_{\rm TRK}\!\approx\!0.2$)~\cite{Harakeh:2001}. As mentioned
earlier, the classical sum rules were derived using a non-relativistic
formalism so one may also need to correct for relativistic 
effects\,\cite{Cohen:2004,Sinky:2006}. We assume here that such 
relativistic effects may also be subsumed into such a correction factor.
However, to our knowledge, a fully relativistic counterpart to 
Eq.\,(\ref{EWSR}) does not exist in the relativistic RPA theory.

In addition to the EWSR a moment of critical importance to the present
work is the {\sl inverse} EWSR. In the non-relativistic formalism, it has
long been established that such a moment can also be computed from 
the ground-state density of the constrained system by invoking the 
``{\sl dielectric theorem}''\,\cite{Bohigas:1978qu,Marshalek:1973}. That is,
\begin{equation}    
  m_{-1}(E0) = \int_{0}^{\infty}\!\omega^{-1} R(\omega;E0)\,d\omega =
  - \frac{1}{2} \left[ \frac{d}{d\lambda} A\langle r^2\rangle_{\lambda}
  \right]_{\lambda = 0} = -\frac{1}{2} \left[ \frac{d}{d\lambda} 
   \int r^{2}\rho({\bf r};\lambda)\,d^{3}r \right]_{\lambda = 0} \;.
 \label{IEWSR}
\end{equation}
The ground-state density in this case must be obtained by supplementing 
the mean-field potential with a ``{\sl constrained}'' one-body term of the 
form $\lambda r^{2}$. The addition of such a term shifts the weight of the 
single-particle orbitals to the interior, thereby leading to a more compact 
system. The inverse EWSR measures the (negative) slope of the 
mean-square radius at $\lambda\!=\!0$. The same constrained procedure 
can be easily implemented in the relativistic approach by adding such a
harmonic one-body potential to the repulsive vector interaction. However, 
in contrast to the simplicity of the prescription, establishing a formal proof 
of the dielectric theorem for a relativistic Hamiltonian is likely to be more 
challenging. Indeed, with due allowance made for the suggestion in Sec.\,3 
of Ref.\,\cite{Ma:2001} that the relativistic RPA $m_{-1}$ can be 
extracted from the constrained RMF ground-state, we have not 
found in the literature a general proof of the validity---or lack thereof---of 
the dielectric theorem within the relativistic theory. Nevertheless, for
our numerical exploration we will follow the non-relativistic approach and
compute the relativistic constrained energy from the $m_{1}$ and $m_{-1}$ 
moments as follows:
\begin{equation}    
  E_{\rm con}\!=\! \sqrt{\frac{m_{1}}{m_{-1}}} \;.
 \label{Econ}
\end{equation}
Note that we have reserved the term {\sl ``centroid energy''} to the
more conventional ratio of $E_{\rm cen}\!=\!m_{1}/m_{0}$. In what
follows we use RMF models to investigate the agreement between 
the constrained energy obtained using the classical sum rules 
[Eqs.(\ref{EWSR}) and (\ref{IEWSR})] and from the corresponding 
integrals derived from the RPA distribution of monopole strength. 
We believe this to be the first RMF study of its kind, except perhaps 
for the study of the $m_{-1}$ moment in ${}^{208}$Pb presented in 
Ref.\,\cite{Ma:2001}. As it will be shown in the next section, we find 
that the relativistic constrained calculation agrees with the corresponding 
RPA value to better than 2\%, suggesting that for many systematic 
applications the constrained calculation may be of enormous utility. 
The result is of obvious practical interest because of the pervasive 
availability of computer codes for RMF spherical ground-state calculations 
as opposed to the scarcity of relativistic RPA codes. Moreover, 
self-consistent relativistic RPA calculations are numerically expensive 
and not free of technical subtleties; whereas they are close to prohibitive 
for large-scale investigations, constrained RMF calculations are simple 
and very fast.

\section{Results}
\label{Results}

\begin{figure}[ht]
\vspace{-0.05in}
\includegraphics[width=0.45\columnwidth,angle=0]{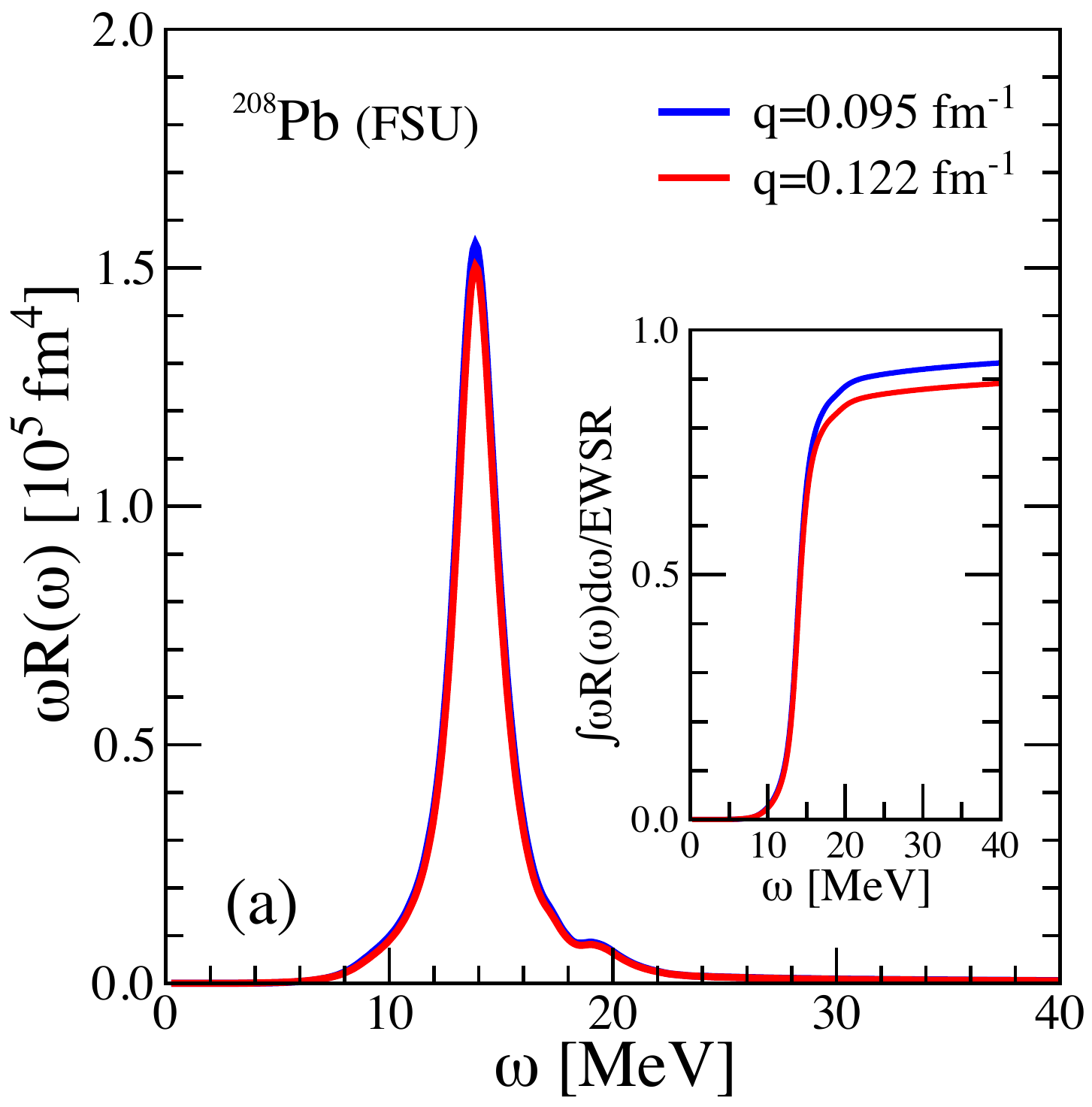}
\includegraphics[width=0.45\columnwidth,angle=0]{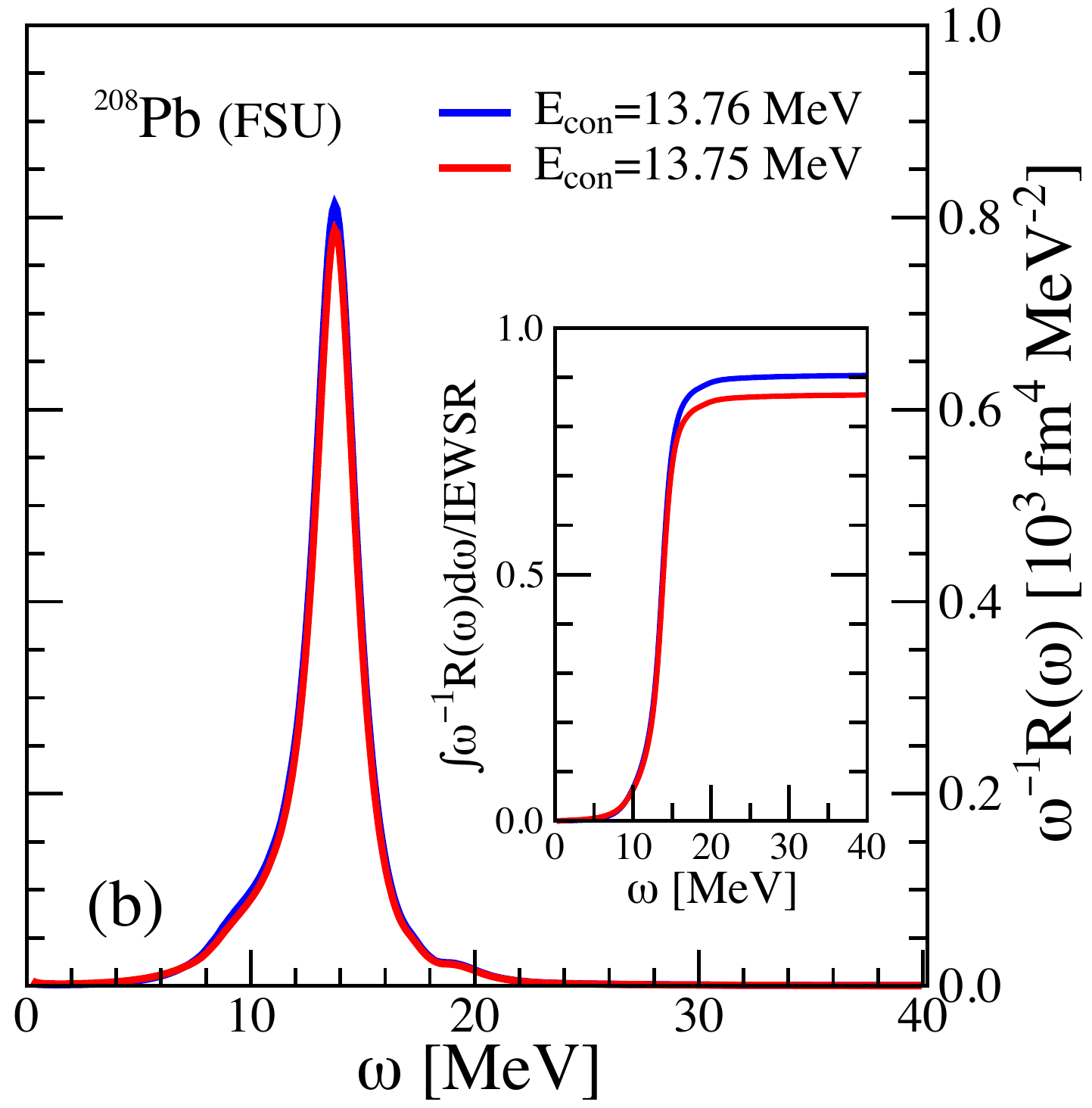}
\caption{(Color online) The energy weighted monopole strength 
  (a) and the {\sl inverse} energy weighted monopole strength (b) 
  in ${}^{208}$Pb for two (small) values of the momentum transfer.
  The insets display the cumulative sums relative to the corresponding
  sum rules obtained from the constrained RMF approach [see
  Eqs.~(\ref{EWSR}) and~(\ref{IEWSR})].}
\label{Fig1}
\end{figure}

We start the section by displaying in Fig.\,\ref{Fig1} the distribution
of isoscalar monopole strength for ${}^{208}$Pb using the accurately
calibrated FSUGold (``FSU'') parametrization~\cite{Todd-Rutel:2005fa}.
Model parameters for this and the NL3 model are listed in
Table~\ref{Table1}.  Note that the left-hand figure displays the
energy weighted monopole strength whereas the right-hand panel the
{\sl inverse} energy weighted strength. We should also mention that
due to the {\sl non-spectral} character of the RPA
approach~\cite{Piekarewicz:2001nm}, the particle-escape width is
computed exactly within the model. Given that the distribution of
monopole strength is defined as the long-wavelength limit of the
longitudinal response [see Eq.~(\ref{RGMR})], we display the strength
function for two small values of the momentum transfer to ensure the
convergence of our results. Finally, the two insets display the
cumulative sums relative to their corresponding sum rules computed
from the constrained RMF approach as indicated in Eqs.~(\ref{EWSR})
and~(\ref{IEWSR}). The insets indicate that the RPA response accounts
for about 90\% of the corresponding sum rules. However, the
constrained energies agree to better than 2\%, namely, 
$E_{\rm con}({\rm CRMF})\!=\!13.50$\,MeV and 
$E_{\rm con}({\rm RPA})\!=\!13.76$\,MeV.  Note, however, that this 2\% 
difference represents the largest discrepancy obtained in the present
work (see Table~\ref{Table2}).
\begin{widetext}
\begin{center}
\begin{table}[h]
\begin{tabular}{|l||c|c|c|c|c|c|c|c|c|c|}
 \hline
 Model & $m_{\rm s}$  & $m_{\rm v}$  & $m_{\rho}$  
       & $g_{\rm s}^2$ & $g_{\rm v}^2$ & $g_{\rho}^2$
       & $\kappa$ & $\lambda$ & $\zeta$ & $\Lambda_{\rm v}$\\
 \hline
 \hline
 NL3       & 508.194 & 782.501 & 763.000 & 104.3871 & 165.5854 &  79.6000 
               & 3.8599  & $-$0.015905 & 0.00 & 0.000 \\
 FSU        & 491.500 & 782.500 & 763.000 & 112.1996 & 204.5469 & 138.4701 
               & 1.4203  & $+$0.023762 & 0.06 & 0.030 \\
\hline
\end{tabular}
\caption{Parameter sets for the two accurately calibrated relativistic
               mean-field models used in the text:
               NL3~\cite{Lalazissis:1996rd,Lalazissis:1999}
               and FSUGold~\cite{Todd-Rutel:2005fa}. The parameter
               $\kappa$ and the meson masses $m_{\rm s}$, $m_{\rm v}$, 
               and $m_{\rho}$ are all given in MeV. The nucleon mass
               has been fixed at  $M\!=\!939$~MeV in both models.}
\label{Table1}
\end{table}
\end{center}
\end{widetext}
Before proceeding any further with the discussion of our results we
should mention some subtleties associated with the extraction of the
constrained energy from the RPA results. Essentially, the potential
problems emerge from the lack of convergence of the integrals over the
excitation energy $\omega$ defining both $m_{1}$ and $m_{-1}$. Indeed, the
lack of convergence of the energy-weighted and inverse energy-weighted sums can be clearly
appreciated in the insets of Figs.~\ref{Fig1}(a) and ~\ref{Fig1}(b), respectively. Note that this
situation is not exclusive to the monopole resonance but extends to
all excitation modes. To elucidate the problem we resort to an ideal
Lorentzian distribution of strength that provides an accurate
representation of the monopole strength in heavy nuclei, such as
${}^{208}$Pb. The Lorentzian distribution is defined as follows:
\begin{equation}    
  R(\omega) = m_{0} \frac{\Gamma/2\pi}
  {(\omega-\omega_{0})^2+\Gamma^{2}/4} \;, 
 \label{RLorentz}
\end{equation}
where $m_{0}$ is the unweighted-energy sum and $\Gamma$ is the full
width at half maximum. Note that the full unweighted-energy sum
$m_{0}$ can only be recovered by extending the integral over the
unphysical $\omega\!<\!0$ region. That is,
\begin{equation}    
  \int_{-\infty}^{\infty}R(\omega)\,d\omega = m_{0} \;.
 \label{m0Lorentz}
\end{equation}
However, by limiting the integral to the physical $\omega\!>\!0$
region, one may still account for most of the unweighted-energy 
sum---especially in the case that $\Gamma\!\ll\!\omega_{0}$.
Indeed, 
\begin{equation}    
  \int_{0}^{\infty}R(\omega)\,d\omega = 
  m_{0}\left[ \frac{1}{2} + \frac{1}{\pi}
  \arctan\left(\frac{\omega_{0}}{\Gamma/2}\right)\right]=
  m_{0}\left[ 1-
  \frac{1}{\pi}\left(\frac{\Gamma/2}{\omega_{0}}\right)+
  \ldots \right]\;.
  \label{m0Lorentz2}
\end{equation}
In the particular case of ${}^{208}$Pb with the parameters predicted
by the FSUGold parametrization ($\omega_{0}\!=\!13.82$\,MeV and
$\Gamma\!=\!2.29$\,MeV) one can still account for about 97\% of the
unweighted-energy sum. However, the situation is radically different
with the $m_{1}$ and $m_{-1}$ moments as both integrals diverge;
$m_{1}$ displays an ultraviolet divergence whereas $m_{-1}$ an
infrared divergence---both logarithmic. Yet both integrals are well
behaved if the integration region is allowed to be extended to the 
unphysical region. That is,
\begin{subequations}    
\begin{eqnarray}
 m_{1} &\equiv& \int_{-\infty}^{\infty}\!\omega 
 R(\omega) d\omega = m_{0}\,\omega_{0} \;, \\
 m_{-1} &\equiv& \int_{-\infty}^{\infty}\!\omega^{-1} 
 R(\omega) d\omega = 
\frac{m_{0}\,\omega_{0}}{\omega_{0}^{2}+\Gamma^{2}/4} \;.
\end{eqnarray}
\label{Moments2}
\end{subequations}
Note that in the case of an RPA distribution of an exact
Lorentzian shape the centroid and constrained energies will be 
given by the following simple expressions:
\begin{equation}    
 E_{\rm cen}({\rm RPA})= \omega_{0} 
 \quad{\rm and}\quad
 E_{\rm con}({\rm RPA})=
 \sqrt{\omega_{0}^{2}+\Gamma^{2}/4} \;.
  \label{EcenEcon}
\end{equation}
So how does one extract the constrained energy from a distribution of
strength that is physically meaningful only for positive values of
$\omega$? Given the previous discussion, we suggest to rely heavily on
the unweighted distribution of strength as it is both well behaved and
accounts for almost 100\% of the sum rule (at least in the Lorentzian
approximation). Thus, we select the upper limit of integration
($\omega_{\rm max}$) in such a way that most of the integrated
strength $m_{0}$ has been accounted for. For the case of the
energy-weighted sum, the remaining contribution (from 
$\omega_{\rm max}$ to $\infty$) is assumed to be exactly cancelled by the
contribution in the unphysical region. Finally, to remove the 
$\omega\!=\!0$ singularity in $m_{-1}$, we replace the 
low-energy tail by a rapidly falling exponential distribution
in the interval $0\!<\omega\!<\omega_{\rm min}$. 
Given that the divergence at both low and high excitation energy 
is logarithmic in nature, we found our results stable against small 
changes in both $\omega_{\rm min}$ and $\omega_{\rm max}$.
  \begin{table}[h]
  \begin{tabular}{|c||c|c|c|c||c|c|c|}
    \hline
    Nucleus & $\omega_{\rm min}$-$\omega_{\rm max}$  
                  & $E_{\rm con}({\rm CRMF})$ & $E_{\rm con}({\rm RPA})$ 
                  & $\Delta E_{\rm con}(\%)$ & $\omega_{\rm min}$-$\omega_{\rm max }$  
                  & $E_{\rm cen}({\rm RPA})$ & $E_{\rm cen}({\rm Exp})$  \\
    \hline
    \hline
     ${}^{16}$O    & 0-50 & 23.34 & 23.35 & 0.04 & 11-40 & 23.95 & 21.13$\pm$0.49  \\
     ${}^{40}$Ca   & 0-50 & 21.55 & 21.57 & 0.09 &10-55 & 21.95 & 19.18$\pm$0.37 \\
     ${}^{90}$Zr    & 0-40 & 18.58 & 18.55 & 0.16 &10-26 & 18.54 & 17.89$\pm$0.20 \\
     ${}^{116}$Sn  & 0-40 & 16.98 & 17.06 & 0.47 &10-23 & 17.03 & 16.07$\pm$0.12 \\
     ${}^{144}$Sm & 0-40 & 16.08 & 16.16 & 0.50 & 10-22 & 16.10 & 15.39$\pm$0.28 \\
     ${}^{208}$Pb  & 0-40 & 14.07 & 14.10 & 0.21 &  8-21 & 14.19 & 14.17$\pm$0.28 \\
    \hline
  \end{tabular}
 \caption{Giant monopole resonance constrained and centroid energies 
                for a variety of nuclei as predicted by the NL3 
                parametrization\,\cite{Lalazissis:1996rd,Lalazissis:1999} .
                Experimental data extracted from
                Refs.\,\cite{Youngblood:1999,Youngblood:2001mq,Lui:2001xh}. 
                All quantities are given in MeV.}
  \label{Table2}
 \end{table}
  \begin{table}[h]
  \begin{tabular}{|c||c|c|c|c||c|c|c|}
    \hline
    Nucleus & $\omega_{\rm min}$-$\omega_{\rm max}$  
                  & $E_{\rm con}({\rm CRMF})$ & $E_{\rm con}({\rm RPA})$
                  & $\Delta E_{\rm con}(\%)$ & $\omega_{\rm min}$-$\omega_{\rm max }$  
                  & $E_{\rm cen}({\rm RPA})$ & $E_{\rm cen}({\rm Exp})$  \\
    \hline
    \hline
     ${}^{16}$O    & 0-50 &  22.89  & 23.09  & 0.87 & 11-40 & 23.68 & 21.13$\pm$0.49  \\
     ${}^{40}$Ca   & 0-50 &  20.67 & 20.66  & 0.05 & 10-55 & 21.18 & 19.18$\pm$0.37 \\
     ${}^{90}$Zr    & 0-40 &  17.70 & 17.94  & 1.34 & 10-26 & 17.88 & 17.89$\pm$0.20 \\
     ${}^{116}$Sn  & 0-40 & 16.20 & 16.48  & 1.70 & 10-23 & 16.47 & 16.07$\pm$0.12 \\
     ${}^{144}$Sm & 0-40 & 15.37 & 15.52  & 0.97 & 10-22 & 15.55 & 15.39$\pm$0.28 \\
     ${}^{208}$Pb  & 0-40 & 13.50 & 13.76  & 1.89 &  8-21 & 13.81 & 14.17$\pm$0.28 \\
    \hline
  \end{tabular}
 \caption{Giant monopole resonance constrained and centroid energies 
                for a variety of nuclei as predicted by the
                FSUGold parametrization\,\cite{Todd-Rutel:2005fa} .
                Experimental data extracted from
                Refs.\,\cite{Youngblood:1999,Youngblood:2001mq,Lui:2001xh}. 
                All quantities are given in MeV.}
  \label{Table3}
 \end{table}

\begin{figure}[ht]
\vspace{-0.05in}
\includegraphics[width=0.6\columnwidth,angle=0]{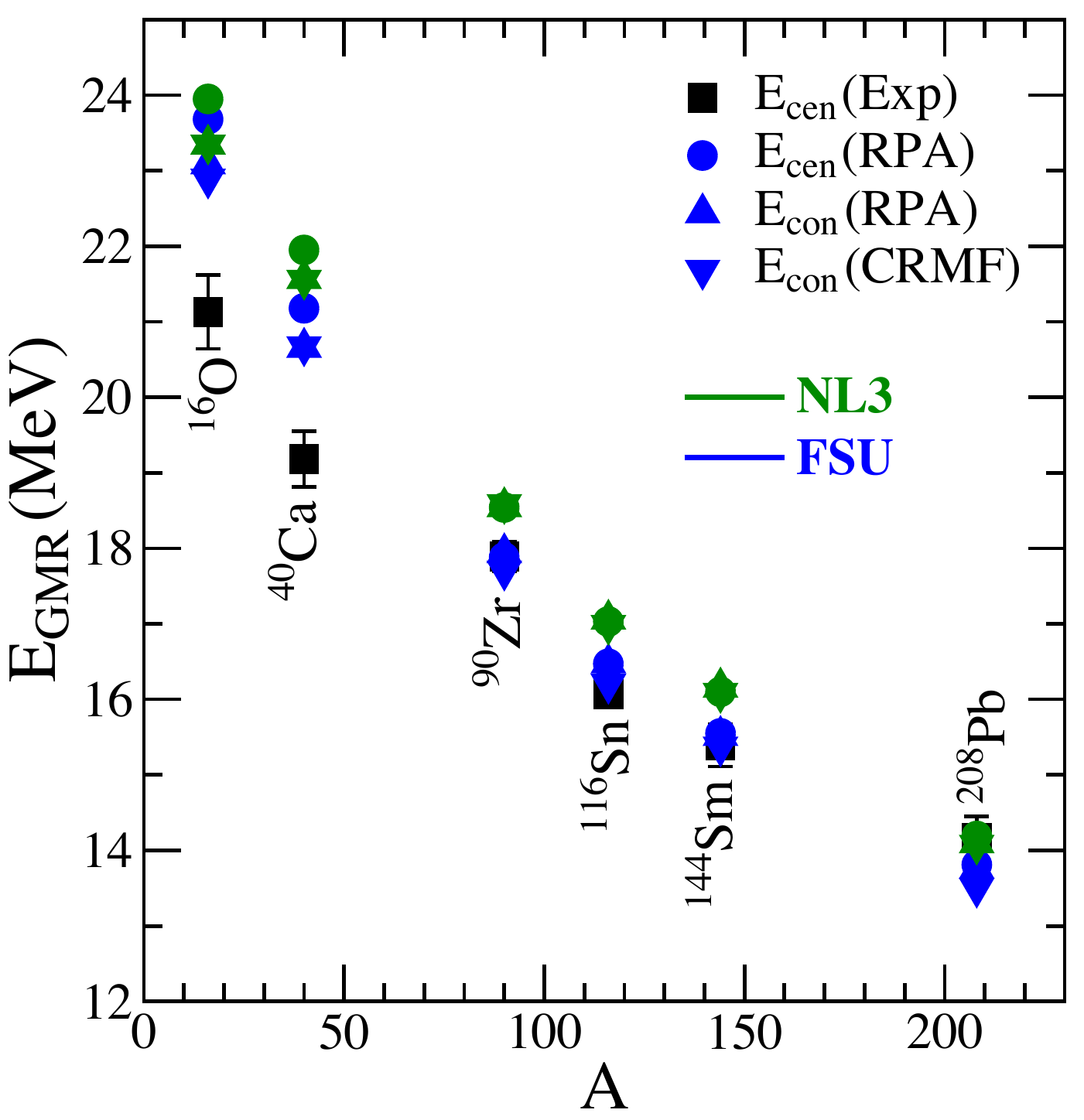}
\caption{(color online) Centroid and constrained energies for a
               variety of nuclei as predicted by the 
               NL3~\cite{Lalazissis:1996rd,Lalazissis:1999} and  
               FSUGold~\cite{Todd-Rutel:2005fa} models. 
               Experimental centroid energies are taken from 
               Refs.~\cite{Youngblood:1999,Youngblood:2001mq,Lui:2001xh}.}
\label{Fig2}
\end{figure}

Having explained how we extract RPA centroid and constrained
energies, we now continue with the discussion of our results. 
Although the use of mean-field methods for light nuclei may be
questionable---especially since the distribution of monopole strength
may be strongly fragmented---we have performed self-consistent RPA
calculations for nuclei ranging from ${}^{16}$O to ${}^{208}$Pb using
the NL3 and FSUGold energy density functionals. Considering these two
functionals is useful because, although they both provide accurate
descriptions of a variety of nuclear ground-state properties, their
predictions for infinite nuclear matter are significantly 
different\,\cite{Piekarewicz:2008nh}. In particular, the incompressibility 
coefficient of symmetric nuclear matter predicted by NL3 is about 271\,MeV 
whereas that of FSUGold is 230\,MeV. We provide our
results in both tabular form in Tables\,\ref{Table2} and \ref{Table3}, 
and graphical form in Fig.\,\ref{Fig2}. First, we note that the NL3 results 
for the centroid energies are systematically higher than the corresponding 
FSUGold results, as expected given the higher incompressibility coefficient 
of the NL3 interaction. The NL3 centroid energies also are systematically 
higher than the experimental data except in the case of ${}^{208}$Pb. 
The case of ${}^{208}$Pb is unique because the large
incompressibility of NL3 is compensated by a similarly large slope of 
the symmetry energy;  this accounts for the accurate prediction of the
centroid energy in ${}^{208}$Pb---but not in 
${}^{90}$Zr\,\cite{Piekarewicz:2003br}.  In essence, the FSUGold
interaction was conceived with the goal of better constraining the
incompressibility of neutron-rich matter by incorporating information
on giant resonances into the calibration procedure. However, in the
present work we aim to go one step further by exploring the feasibility 
of incorporating monopole energies directly into the calibration of the
quality function. The results presented in both of the tables are extremely 
encouraging as they suggest differences of at most 2\% between energies 
computed with the constrained RMF approach and those extracted from 
the RPA strength function. For reasons that at present we do not understand,
the discrepancy between constrained and RPA results is systematically
higher in the case of the FSUGold parametrization. Nevertheless, the
accuracy of our results gives us confidence that in the future, constrained 
GMR energies for nuclei ranging from ${}^{90}$Zr to ${}^{208}$Pb---and
including a variety of Sn-isotopes~\cite{Li:2007bp,
Piekarewicz:2007us,Piekarewicz:2009gb}---may be directly included in
the accurate calibration of energy density functionals.  Such a
procedure will not only be able to better constrain the density
dependence of the equation of state around saturation density, but may
be helpful in partly removing the reliance on {\sl ``pseudo-data''} ({\sl i.e.,}
on bulk properties of infinite nuclear matter).
 
\section{Conclusions}
\label{Conclusions}

The sum-rule approach provides a powerful tool for the analysis of
nuclear excitation spectra\,\cite{Harakeh:2001}. Particularly useful
are the energy-weighted sum rules---with the TRK sum rule as its best
known exponent---as they can be determined largely model-independently
from a few well-known ground-state properties.  Moreover, additional
sum rules, such as the inverse energy weighted sum rule,
may be combined to estimate the mean excitation energy of
the resonances from intrinsic ground-state properties. 
However, the standard analytical proofs of the validity of the
sum-rule theorems rely on a {\sl non-relativistic} Hamiltonian. Such
is the case of the model-independent EWSR based on commutation
relations, and of the inverse EWSR extracted from constrained
calculations of the ground state. Thus, the sum-rule approach to giant 
monopole energies has been limited to the non-relativistic domain;
for a recent and successful implementation of such techniques see
Ref.\,\cite{Sil:2006sh} and references therein. Indeed, in
Ref.\,\cite{Sil:2006sh} highly accurate Skyrme-based mean-field plus
RPA calculations are successfully compared against constrained
predictions for the mean excitation energy of various resonances.

In this contribution we have conjectured that the constrained approach 
to the isoscalar GMR may be extended to the relativistic domain. We 
were prompted by the fact that in different applications, as long as the 
energy density functional is accurately calibrated to physical data, the 
specific form of the nuclear functional becomes a matter of convenience. 
Indeed, accurately calibrated non-relativistic and relativistic energy density 
functionals provide a similar distribution of the isoscalar monopole strength.
We believe that assessing the degree of agreement in the relativistic domain 
between the predictions of the constrained approach and the RPA results 
is reason enough to implement the present study. However, our study also 
has an important practical component.  Given that we aim to eventually 
supplement the calibration of energy density functionals with experimental 
data on nuclear collective modes, the validity of the constrained approach
becomes essential, as it is impractical to incorporate this information from a 
detailed RPA calculation.
 
Relativistic mean-field plus RPA calculations were performed for the
distribution of isoscalar monopole strength for magic or semi-magic
nuclei ranging from ${}^{16}$O to ${}^{208}$Pb using two accurately
calibrated density functionals (NL3 and FSUGold).  Predictions were
made for both centroid and constrained energies from the various
moments of the distribution. RPA calculations of this kind are
computationally expensive as the strength distribution is finely
mapped over a wide energy range.  In contrast, the extraction of
monopole energies from the constrained approach is numerically
expedient as it relies exclusively on the self-consistent evaluation
of ground-state properties. As previously conjectured, we find
excellent agreement between the two approaches. Indeed, for all nuclei
under consideration the discrepancy between the two approaches 
amounts to less than 2\%, a quantity that is comparable to current
experimental errors.

In summary, relativistic constrained calculations of monopole energies were
favorably compared against the corresponding predictions from the 
relativistic RPA approach. These results are highly encouraging and show the 
promise of incorporating nuclear excitations in the calibration of future 
energy density functionals without incurring in an unaffordable computational cost.

\begin{acknowledgments}
M.C. is grateful to Prof. J. Martorell for useful discussions. This work was supported 
in part by the United States Department of Energy under grant DE-FG05-92ER40750. 
M. C. acknowledges support from the Consolider Ingenio 2010 Programme CPAN 
CSD2007-00042, from Grant No. FIS2011-24154 from MICINN and FEDER, and from 
Grant No. 2009SGR-1289 from Generalitat de Catalunya. 
\end{acknowledgments}

\bibliography{ConstrainedRMF.bbl}

\end{document}